\documentclass[footinbib,aps,pre,reprint,showpacs,superscriptaddress,longbibliography]{revtex4-1}
\usepackage{amsmath,amssymb,graphicx,units}
\usepackage[usenames,dvipsnames]{color}

\newcommand{\ipcms}{Universit{\'e} de Strasbourg, CNRS, Institut de Physique et Chimie des Mat{\'e}riaux de Strasbourg, UMR 7504, F-67000 Strasbourg, France}

\begin{document}
\title{Does a Single Eigenstate of a Hamiltonian Encode the Critical Behaviour of its Finite-Temperature Phase Transition?}

\author{Keith R.\ Fratus}
\affiliation{\ipcms}
\author{Syrian V. Truong}
\affiliation{\ipcms}
\affiliation{Department of Physics, University of California, Santa Barbara, California, 93106, USA}
\affiliation{Division of the Physical Sciences, University of Chicago, Illinois, 60637, USA}

\begin{abstract}

Recent work on the subject of isolated quantum thermalization has suggested that an individual energy eigenstate of a non-integrable quantum system may encode a significant amount of information about that system's Hamiltonian. We provide a theoretical argument, along with supporting numerics, that this information includes the critical behaviour of a system with a second-order, finite-temperature phase transition. 

\end{abstract}

\maketitle

\section{Introduction}

The eigenstate thermalization hypothesis (ETH) \cite{deutsch1991quantum, srednicki1994chaos, srednicki95, srednicki99, rigol_dunjko_08} has recently been the subject of a large body of experimental and theoretical work \cite{rigol_09a, rigol_09b, santos_rigol_10b,santos_rigol_10a, neuenhahn_marquardt_12, genway_ho_12, khatami_pupillo_13, beugeling_moessner_14, kim_14, sorg_vidmar_14, AETRigol,FSSIkeda,PLETHSteinigweg,Khlebnikov2013yia,Garrison2015lva,NDPolkovnikov}. ETH can explain how an isolated, quantum many-body system in an initial pure state can come to thermal equilibrium (as determined by measurements of a specified set of observables) in finite time, and is thus fundamental to understanding the validity of conventional quantum statistical mechanics as an accurate description of the long-time behavior of quantum systems; for a review, see \cite{mrigolreview2015}. ETH is expected to hold in systems without disorder that are sufficiently far from integrability 
(including effective integrability caused by many-body localization in disordered systems), 
for observables that are sufficiently simple functions of the fundamental degrees of freedom.

While the key statements of ETH have been cast in several forms by various authors, in this work we will be interested in what we will refer to as the strong version of ETH \cite{Garrison2015lva}, which concerns the entanglement behaviour of individual energy eigenstates in a non-integrable quantum system. This strong version of ETH is the statement that within one individual energy eigenstate of a non-integrable quantum system, the reduced density matrix (RDM) of a sufficiently small subsystem will resemble that of a thermal one. More precisely, the RDM on a subsystem $A$, constructed from an eigenstate $|\psi\rangle$ of the full quantum system, will be approximately given by
\begin{equation}
\rho^{\psi}_{A} \sim \frac{e^{- \hat{H}_{A} / T_{\psi}}}{\mathcal{Z}( \hat{H}_{A},T_{\psi}  )}~;~\mathcal{Z}(\hat{H}_{A},T_{\psi}   ) \equiv \text{Tr} \left [ e^{- \hat{H}_{A} / T_{\psi}} \right ]
\label{eq:redMat}
\end{equation}
where $T_{\psi}$ is the ``temperature'' of the thermal eigenstate $|\psi\rangle$, and $\hat{H}_{A}$ is the Hamiltonian of the subsystem $A$. The heuristic interpretation of this statement is that even in a pure energy eigenstate, the full system acts as a thermal reservoir for its small subsystems, thermalizing them through the quantum entanglement between the subsystem and the larger thermal reservoir.

Previous authors have made precise the exact definitions of $T_{\psi}$ and $\hat{H}_{A}$ which are necessary for the above statement to hold, and have also elucidated the conditions under which it should be expected to hold, and how well \cite{Garrison2015lva,newStanford}. Here we will not be focused with these details, as they have been thoroughly addressed previously. We will, however, be concerned with an interesting corollary to equation \ref{eq:redMat}, mentioned already in \cite{Garrison2015lva}. Given a thermal density matrix for a system at one temperature, it is always possible to compute the density matrix at another temperature, by simply raising the density matrix to a power and re-normalizing it, since
\begin{equation}
\left (\frac{e^{- \hat{H} / T_{1}}}{\mathcal{Z}  ( \hat{H},T_{1}  )}  \right )^{T_{1} / T_{2}} = ~\frac{e^{- \hat{H} / T_{2}}}{\left [\mathcal{Z}  ( \hat{H},T_{1}  )  \right ]^{T_{1} / T_{2}}}.
\label{eq:extrap}
\end{equation}
Thus, in principle, the RDM of a small subsystem, extracted from a thermal eigenstate of a much larger non-integrable quantum system, should posses information about the thermal behaviour of this subsystem across a range of temperature scales, to the extent that equation \ref{eq:redMat} is a valid approximation. This claim has in fact been investigated thoroughly by previous authors, and has been verified to be true under appropriate circumstances, outlined in \cite{Garrison2015lva}.

Here we will focus on what this information reveals about the behaviour of a quantum, non-integrable system with a second-order phase transition at finite temperature. Previous work \cite{myPaperOne,myPaperTwo,myPaperThree} has suggested that ETH should also be expected to hold in such systems, and in fact signatures of ETH and quantum chaos have indeed been found to exist even in the broken symmetry phase of such a system. In the present paper, we will argue that if such a system with a finite-temperature phase transition satisfies the strong version of ETH, then individual energy eigenstates of this system can diagnose the existence of this phase transition, and will also contain quantitative information about its critical behaviour, without any knowledge of the original Hamiltonian itself. Below, we outline a procedure by which one could arrive at this information from such an energy eigenstate.

\section{The Extraction Procedure}

Consider a non-integrable quantum system, with some Hamiltonian $\hat{H}$, which may or may not be known to us. We suspect that this system may posses a second-order phase transition at finite temperature, with corresponding order parameter $\mathcal{Q}$, and we wish to diagnose this fact from the information contained in one individual eigenstate of this Hamiltonian. We will assume that a reasonable notion of ``subsystem'' can be defined in our system. We will also assume that the system, while finite, is sufficiently ``large,'' such that subsystems can be defined which are themselves ``large,'' yet still much smaller than the full system. For example, in a spin system consisting of $N$ spins, we imagine that $N$ is large enough such that we can define subsystems of $n$ spins, with $1 << n << N$.

We now consider a single energy eigenstate $|\psi\rangle$ of this quantum system. Under the assumption that this eigenstate is thermal, with some characteristic temperature $T_{\psi}$, the RDMs constructed on a sufficiently small subsystem should be appropriately thermal, in the sense described by equation \ref{eq:redMat}.  Based upon the arguments outlined in \cite{Garrison2015lva}, we should be able to probe the thermal density matrix of such a subsystem across a wide range of temperatures,
\begin{equation}
\widetilde{\rho}_{A} \left (T  \right ) \equiv \left ( \rho_{A} \right )^{1/T} \sim \left (e^{- \hat{H}_{A} / T_{\psi}}  \right ) ^{1/T} = e^{- \widetilde{H}_{A} / T},
\label{eq:relextrap}
\end{equation}
where we have defined
\begin{equation}
\widetilde{H}_{A} \equiv \hat{H}_{A} / T_{\psi}
\label{eq:scaleHam}
\end{equation}
to be the ``scaled,'' dimensionless subsystem Hamiltonian, and $\widetilde{\rho}_{A} \left (T  \right )$ its canonical density matrix at temperature $T$. Without knowledge of the original Hamiltonian, it is not possible to determine this $T_{\psi}$, and thus the absolute temperature scale of our predictions, but this information will not be necessary for our purposes; we will simply assume that such a $T_{\psi}$ exists, and thus make thermal predictions regarding the scaled Hamiltonian. Such an overall scale factor will not affect any predictions regarding the critical exponents of such a Hamiltonian. 

With such a thermal density matrix at an arbitrary temperature, the thermal expectation value of any observable $\mathcal{O}_{A}$ which lives on this subsystem can be computed at this temperature,
\begin{equation}
\langle \mathcal{O}_{A} \rangle = \text{Tr} \left [ \widetilde{\rho}_{A} \mathcal{O}_{A} \right ].
\label{eq:expect}
\end{equation}
In particular, the observable in question could be the order parameter of the system, its associated susceptibility, or any relevant correlation functions. Since this procedure could be repeated for various subsystems of different sizes, it should be possible to perform a finite-size scaling analysis of these quantities as a function of temperature, thus allowing for the quantitative extraction of various critical exponents \cite{cardy2012finite}.

Any such finite-size scaling analysis will of course be limited by the size of the system $N$, but if we anticipate that equation \ref{eq:redMat} holds for arbitrarily large system sizes, then one can always extract these critical exponents to the desired accuracy by considering sufficiently large $N$. Thus, as we approach the thermodynamic limit, the information about the critical point extracted from a single eigenstate can be made arbitrarily accurate, and it is in this limit in which a single eigenstate will encode the full information regarding the critical point of the original Hamiltonian. 

To see how this procedure could work in somewhat more detail, we provide a specific example involving a system of $N$ Ising spins defined on a lattice, and examine the order parameter which is the total magnetization in the z-direction,
\begin{equation}
\mathcal{Q} = \hat{M}_{z} \equiv \sum_{i} \hat{\sigma}_{i}^{z}.
\label{eq:spinOrder}
\end{equation}
This order parameter would be the relevant one for systems possessing an Ising transition. As a result of the Ising symmetry, any \textit{finite} system will always posses
\begin{equation}
\langle \hat{M}_{z} \rangle = 0,
\label{eq:finitespinOrder}
\end{equation}
and so a more useful metric for studying the critical behaviour of this model is given by the Binder cumulant of the order parameter,
\begin{equation}
U \equiv 1 - \frac{\langle \hat{M}_{z}^{4} \rangle}{3 \langle \hat{M}_{z}^{2} \rangle^{2}},
 \label{eq:binder}
\end{equation}
At low temperatures, in a system with an Ising transition, the Binder cumulant approaches a value of $2/3$, up to corrections which scale as $1/N$, while at high temperatures it approaches a value of zero, again up to corrections which scale as $1/N$. In the large system size limit, the transition between these two Binder cumulant values is sharp, transitioning between the two limiting cases at the critical temperature of the model, $T_{c}$. When the Binder cumulant is plotted as a function of temperature for different finite system sizes, the crossing point of these curves provides a good estimate for the critical temperature, $T_{c}$ \cite{Binder81}.

If we now imagine starting from one energy eigenstate of this system, computing the RDM for many different subsystems $\{A\}$, all satisfying $1 << n_{A} << N$, and then using these RDMs to study the behaviour of the Binder cumulant as a function of temperature for all of these different subsystem sizes, we can diagnose the existence of such an Ising transition by demonstrating the existence of a crossing point at some non-zero $T_{c}$.

Furthermore, if we wish to extract quantitative information about the critical behaviour of this transition, we could, for example, extend this procedure to the magnetic susceptibility, which, for a finite system of size $n_{A}$, reaches a maximum at a pseudo-critical point, $T_{c} \left (n_{A} \right )$. Combining this fact with our RDM procedure to extract $T_{c} \left (n_{A} \right )$ for many different subsytem sizes, and then using the scaling relation \cite{cardy2012finite}
\begin{equation}
T_{c}^{-1} \left ( n_{A} \right ) = T_{c}^{-1} \left ( \infty \right ) - a n_{A}^{-1/\nu},
\label{eq:scaling}
\end{equation}
where $a$ is some constant and $T_{c}^{-1} \left ( \infty \right )$ is the (inverse) critical temperature in the thermodynamic limit, it becomes possible to extract the critical exponent $\nu$ describing the divergence of the spin-spin correlation length. Since this analysis was performed using RDMs that were extracted from one individual energy eigenstate, this critical exponent must have been encoded in this state. A similar analysis could be performed for any other critical exponent of interest.

We mention here two important subtleties of this procedure, and argue why they should not substantially alter any of the conclusions reached above. First, there is some subtlety that is associated with the proper definition of $\hat{H}_{A}$, in particular, how the matter of boundary terms between the subsystem and its complement should be addressed. Here, when we must make such a distinction, we will adopt the simple (yet possibly less appropriate, see \cite{Garrison2015lva}) convention that $\hat{H}_{A}$ consists of all terms in the original Hamiltonian $\hat{H}$ with support on subsystem $A$, and disregard any terms which involve operators with support on the complementary region, or the boundary between the two. While the precise definition of $\hat{H}_{A}$ will have consequences for the satisfaction of equation \ref{eq:redMat} in a finite system, the distinction is expected to become irrelevant in the thermodynamic limit. Furthermore, if our interest is in merely extracting information about the critical point, our only concern is that the $\hat{H}_{A}$ we recover possesses a finite-temperature phase transition which is in the same universality class as the original Hamiltonian, which is a much weaker requirement. 

Second, we emphasize that in our finite-size scaling analysis, \textit{all observable quantities, as a function of temperature, should be analytic.} One may object to the procedure described above, in that we are taking a thermal density matrix which may correspond to a temperature on one side of the critical point of our model, and using it to extrapolate the behaviour of the subsystem to temperatures which are on the other side of the critical point. However, since there is no actual phase transition in a strictly finite system, there is no concern that we are attempting to extrapolate an observable quantity across a singularity.

\section{Numerical Investigation}

Our theoretical argument outlined above rests on the assumption that the strong form of ETH will be satisfied in quantum systems with a finite-temperature phase transition, and that it is indeed possible to use one energy eigenstate of such a system to extrapolate observable quantities for a subsystem across a wide range of temperatures. We now provide numerical evidence in support of this claim. Our model Hamiltonian will be the transverse-field Ising chain
\begin{equation}
 \hat H = -\sum_{i \neq j}J_{ij}\hat\sigma_{ i}^z\hat\sigma_{ j}^z - g\sum_{ i}\hat\sigma_{ i}^x ,
 \label{eq:hamiltonian}
\end{equation}
where $\hat{\sigma}_{ i}^z$ and $\hat{\sigma}_{ i}^x$ are the standard Pauli matrices on site $i$ of a one-dimensional lattice. The Ising interaction $J_{ij}$ is chosen to obey a power-law decay,\begin{equation}
 J_{ij} = \frac{J}{|i-j|^{p}}.
\end{equation}
We set $J=1$, which fixes the energy scale, and corresponds to a ferromagnetic Ising coupling. For the transverse term, we choose $g = 1.5$. Our boundary conditions are chosen to be open, so we do not make use of translation symmetry in diagonalizing the Hamiltonian. We do, however, make explicit use of spatial parity symmetry and Ising symmetry. In our work, we choose $N = 23$, as this is the largest system size for which we are able to find a significant number of exact energy eigenstates. We note that while an individual energy eigenstate of the 23-site system lives within a single sector of the Ising and spatial parity symmetries, the RDMs which we will extract from these eigenstates live on the full Hilbert space of the subsystem they describe. It is a straight-forward exercise to verify that such an RDM will obey the same symmetries as the subsystem Hamiltonian, so long as the original state it is extracted from is an eigenstate of the corresponding symmetries of the 23-site Hamiltonian.

Previous work \cite{myPaperThree} has studied the compatibility between spontaneous symmetry breaking and ETH in this model, for the case that $p = 1.5$, in which there is a second-order phase transition at finite temperature. Here we shall study this case, as well as the case $p = 3.0$, for which there is no finite-temperature phase transition. Both models possess an energy which scales extensively with system size, and thus a well-defined thermodynamic limit \cite{Dyson1969}. For the small system sizes we are able to study numerically, we find that ETH is best satisfied in the Ising chain, as opposed to the model defined on a square lattice, hence the reason for our particular choice of geometry; see \cite{myPaperThree} for details. We will not be interested in any quantum phase transitions which may occur at zero temperature as a result of adjusting any parameters in the Hamiltonian.

\begin{figure}[h]
\centering
\includegraphics[width=80mm]{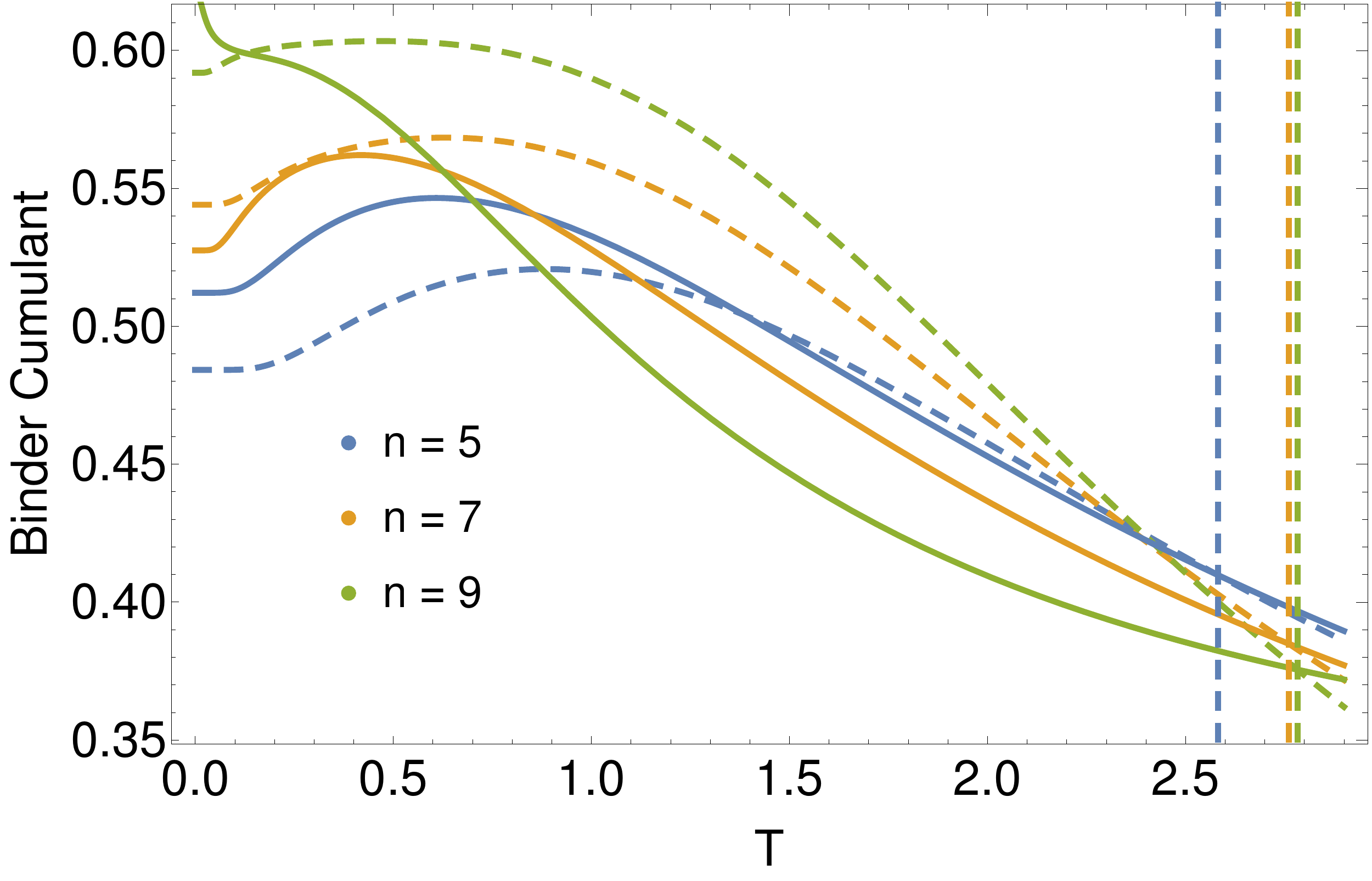}
\caption{The Binder cumulant of the $p = 1.5$ model as a function of temperature for several subsystem sizes. We display both the exact prediction, using the exact subsystem Hamiltonian to compute the canonical partition function (dashed curve), as well as the prediction found from the reduced density matrix procedure (solid curve). The RDMs in this case are extracted from the $3,950^{\text{th}}$ excited state of the even Ising, even parity symmetry sector, which corresponds to an energy of $E = -26.479$. The $T_{\psi}$ associated with each subsystem size are denoted by the vertical dashed lines, which correspond to $T_{\psi} = 2.582$, $2.761$, and $2.783$ for the $n = 5$, $7$, and $9$ subsystems, respectively.}
\label{fig:BCvT15}
\end{figure}

\begin{figure}[h]
\centering
\includegraphics[width=80mm]{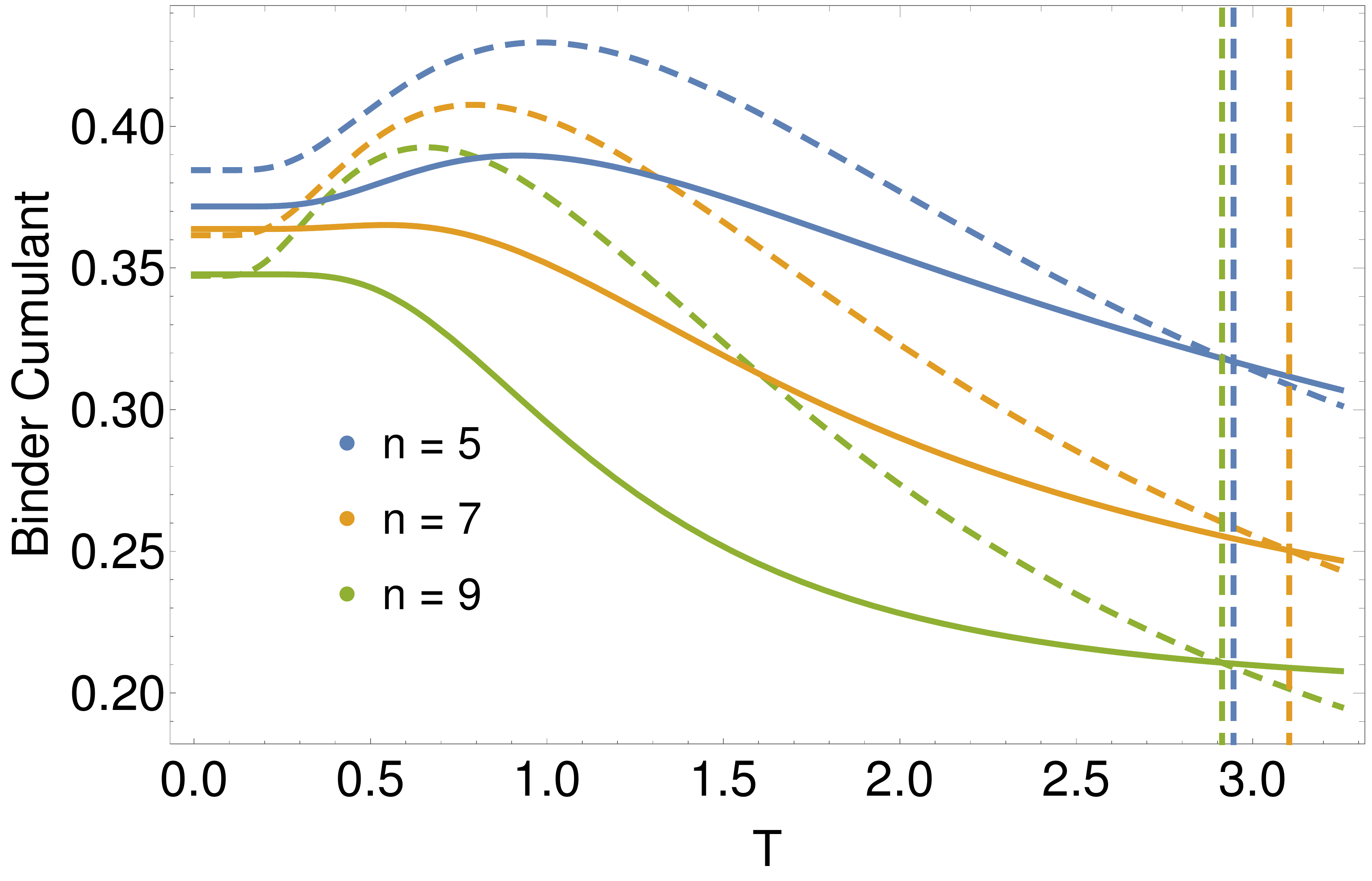}
\caption{The Binder cumulant of the $p = 3.0$ model as a function of temperature for several subsystem sizes. We display both the exact prediction, using the exact subsystem Hamiltonian to compute the canonical partition function (dashed curve), as well as the prediction found from the reduced density matrix procedure (solid curve). The RDMs in this case are extracted from the $3,986^{\text{th}}$ excited state of the even Ising, even parity symmetry sector, which corresponds to an energy of $E = -24.519$. The $T_{\psi}$ associated with each subsystem size are denoted by the vertical dashed lines, which correspond to $T_{\psi} = 2.946$, $3.103$, and $2.913$ for the $n = 5$, $7$, and $9$ subsystems, respectively.}
\label{fig:BCvT3}
\end{figure}

Figures \ref{fig:BCvT15} and \ref{fig:BCvT3} show the result of using this procedure to extrapolate the Binder cumulant as a function of temperature, for the $p = 1.5$ and $p = 3.0$ models, respectively. We perform this procedure for various different subsystem sizes, always with the RDM extracted from the center of the full system, and always from the same energy eigenstate. We also display a comparison against the results obtained from using the exact subsystem Hamiltonian to compute the thermal density matrix of the subsystem directly. While the results of this procedure vary slightly from state to state \footnote{See Supplemental Material at [URL will be inserted by publisher] for a more detailed explanation of the precise numerical extraction procedure we have used, which includes a discussion of how well the subsystem Hamiltonian agrees with the exact Hamiltonian, how we define the temperature of an eigenstate, and how well our results agree from one choice of eigenstate to another.}, we find that most states produce qualitatively similar results, and so we choose to focus here on the $3,950^{\text{th}}$ excited state of the even Ising, even parity symmetry sector for the $p = 1.5$ model, and the $3,986^{\text{th}}$ excited state of the even Ising, even parity symmetry sector for the $p = 3.0$ model.

In order to aide in the visual comparison of these results, we have rescaled the predictions from the RDM method by the temperature $T_{\psi}$ of these eigenstates, so that they appear on the same horizontal scale as the predictions of the exact Hamiltonian; however, we again emphasize that the knowledge of this $T_{\psi}$ is not necessary for the hypothetical extraction procedure we have outlined in the previous section. For the small system sizes considered here, we find that $T_{\psi}$ depends slightly on the choice of subsystem \cite{Note1}.

We note that in both models, given an eigenstate with $T_{\psi} > 0$, all of the extrapolated curves approach their correct $T=0$ values (close to $2/3$ for the $p=1.5$ model, and significantly less than $2/3$ for the $p=3.0$ model). In the case of the $p = 1.5$ model, there is a crossing at some intermediate temperature, \textbf{indicating the presence of a phase transition}, while no such crossing exists for the $p = 3.0$ model. In both cases, the scaling of the Binder cumulant with system size is correct. At sufficiently high temperature, this corresponds to a decreasing Binder cumulant with increasing system size for both models, while at low temperature, this scaling behaviour is reversed below the crossing point for the $p=1.5$ model only. While the quantitative agreement here is not perfect (for example, the precise location of the crossing point for the $p=1.5$ model is not correct), the qualitative agreement is still impressive, given our somewhat simple definition of the subsystem Hamiltonian which neglects the subtle issue of boundary terms, as well as the extremely small system sizes we have been restricted to, due to computational limitations.

We note that, as described in \cite{Garrison2015lva}, the range of temperatures over which we can expect this procedure to work is limited to those in which the total energy of the subsystem does not exceed the energy of the original eigenstate it was extracted from. While this restriction is not a concern when $n << N$, it is relevant when the subsystem in question is a substantial fraction of the full system, which is the case for our numerics displayed here. We have carefully verified that this condition is not violated for the range of temperatures which we have chosen to display in these figures \cite{Note1}.

We do not attempt to perform an actual finite-size scaling analysis of the critical exponents here, as we are limited to system sizes small enough that such a scaling analysis would not yield useful results. However, our numerics still provide strong qualitative evidence that such a procedure, given the eigenstate of a sufficiently large system, should be feasible, so long as the success of the extrapolation procedure we have seen here continues to improve with larger system sizes.

\section{Concluding Remarks}

We have provided a theoretical argument in support of the claim that individual energy eigenstates of a non-integrable Hamiltonian should encode information about whether that Hamiltonian possesses a finite-temperature phase transition, as well as quantitative information about the critical behaviour of this transition. We have also provided numerical evidence in support of this claim, though we have stopped short of actually extracting any quantitative information about such a transition, due to computational limitations. However, under the assumption that the extrapolation procedure we have outlined works for larger system sizes, the existence of such a procedure suggests that individual energy eigenstates, even at energy densities away from the critical point, should contain information about the finite-temperature phase transition of the Hamiltonian they originate from. Such a conclusion may serve as an inspiration for the development of new computational techniques aimed at isolating the critical information contained in one energy eigenstate, without the need to perform a computationally costly exact diagonalization.

While this result strikes the authors as being primarily of interest for questions regarding the foundational principles of quantum statistical mechanics, we note one hypothetical scenario in which it could be of practical use. We imagine a Hamiltonian which is believed to satisfy ETH, and also possesses a sign problem, so that it is difficult to study in a traditional quantum Monte Carlo approach. If, for some reason, there existed a special ansatz that allowed for one to find some small fraction of the spectrum (perhaps, for example, the ground state and first excited state), in such a way that RDMs could be extracted and manipulated in a computationally tractable fashion, then our approach outlined here could be useful. This could have applications, for example, in the study of high-temperature superconductivity. However, at present, the authors do not possess any knowledge of such a hypothetical Hamiltonian or corresponding ansatz.

\begin{acknowledgments}

We thank Mark Srednicki, Dietmar Weinmann, Rodolfo Jalabert, James Garrison, Tarun Grover, and Johannes Schachenmayer for helpful discussions. This work was supported in part by the UCSB McNair Scholars Program, Edison International, and the French National Research Agency ANR through project Grant No. ANR-14-CE36-0007-01 (SGM-Bal). We acknowledge support from the Center for Scientific Computing from the CNSI, MRL: an NSF MRSEC (DMR-1720256) and NSF CNS-1725797.

\end{acknowledgments}

\bibliography{CEE}

\begin{thebibliography}{30}%
\makeatletter
\providecommand \@ifxundefined [1]{%
 \@ifx{#1\undefined}
}%
\providecommand \@ifnum [1]{%
 \ifnum #1\expandafter \@firstoftwo
 \else \expandafter \@secondoftwo
 \fi
}%
\providecommand \@ifx [1]{%
 \ifx #1\expandafter \@firstoftwo
 \else \expandafter \@secondoftwo
 \fi
}%
\providecommand \natexlab [1]{#1}%
\providecommand \enquote  [1]{``#1''}%
\providecommand \bibnamefont  [1]{#1}%
\providecommand \bibfnamefont [1]{#1}%
\providecommand \citenamefont [1]{#1}%
\providecommand \href@noop [0]{\@secondoftwo}%
\providecommand \href [0]{\begingroup \@sanitize@url \@href}%
\providecommand \@href[1]{\@@startlink{#1}\@@href}%
\providecommand \@@href[1]{\endgroup#1\@@endlink}%
\providecommand \@sanitize@url [0]{\catcode `\\12\catcode `\$12\catcode
  `\&12\catcode `\#12\catcode `\^12\catcode `\_12\catcode `\%12\relax}%
\providecommand \@@startlink[1]{}%
\providecommand \@@endlink[0]{}%
\providecommand \url  [0]{\begingroup\@sanitize@url \@url }%
\providecommand \@url [1]{\endgroup\@href {#1}{\urlprefix }}%
\providecommand \urlprefix  [0]{URL }%
\providecommand \Eprint [0]{\href }%
\providecommand \doibase [0]{http://dx.doi.org/}%
\providecommand \selectlanguage [0]{\@gobble}%
\providecommand \bibinfo  [0]{\@secondoftwo}%
\providecommand \bibfield  [0]{\@secondoftwo}%
\providecommand \translation [1]{[#1]}%
\providecommand \BibitemOpen [0]{}%
\providecommand \bibitemStop [0]{}%
\providecommand \bibitemNoStop [0]{.\EOS\space}%
\providecommand \EOS [0]{\spacefactor3000\relax}%
\providecommand \BibitemShut  [1]{\csname bibitem#1\endcsname}%
\let\auto@bib@innerbib\@empty
\bibitem [{\citenamefont {Deutsch}(1991)}]{deutsch1991quantum}%
  \BibitemOpen
  \bibfield  {author} {\bibinfo {author} {\bibfnamefont {J.~M.}\ \bibnamefont
  {Deutsch}},\ }\bibfield  {title} {\enquote {\bibinfo {title} {Quantum
  statistical mechanics in a closed system},}\ }\href@noop {} {\bibfield
  {journal} {\bibinfo  {journal} {Phys. Rev. A}\ }\textbf {\bibinfo {volume}
  {43}},\ \bibinfo {pages} {2046} (\bibinfo {year} {1991})}\BibitemShut
  {NoStop}%
\bibitem [{\citenamefont {Srednicki}(1994)}]{srednicki1994chaos}%
  \BibitemOpen
  \bibfield  {author} {\bibinfo {author} {\bibfnamefont {M.}~\bibnamefont
  {Srednicki}},\ }\bibfield  {title} {\enquote {\bibinfo {title} {Chaos and
  quantum thermalization},}\ }\href@noop {} {\bibfield  {journal} {\bibinfo
  {journal} {Phys. Rev. E}\ }\textbf {\bibinfo {volume} {50}},\ \bibinfo
  {pages} {888} (\bibinfo {year} {1994})}\BibitemShut {NoStop}%
\bibitem [{\citenamefont {Srednicki}(1995)}]{srednicki95}%
  \BibitemOpen
  \bibfield  {author} {\bibinfo {author} {\bibfnamefont {M.}~\bibnamefont
  {Srednicki}},\ }\bibfield  {title} {\enquote {\bibinfo {title} {Thermal
  fluctuations in quantized chaotic systems},}\ }\href@noop {} {\bibfield
  {journal} {\bibinfo  {journal} {Journal of Physics A: Mathematical and
  General}\ }\textbf {\bibinfo {volume} {29}},\ \bibinfo {pages} {L75}
  (\bibinfo {year} {1995})}\BibitemShut {NoStop}%
\bibitem [{\citenamefont {Srednicki}(1999)}]{srednicki99}%
  \BibitemOpen
  \bibfield  {author} {\bibinfo {author} {\bibfnamefont {M.}~\bibnamefont
  {Srednicki}},\ }\bibfield  {title} {\enquote {\bibinfo {title} {The approach
  to thermal equilibrium in quantized chaotic systems},}\ }\href@noop {}
  {\bibfield  {journal} {\bibinfo  {journal} {Journal of Physics A:
  Mathematical and General}\ }\textbf {\bibinfo {volume} {32}},\ \bibinfo
  {pages} {1163} (\bibinfo {year} {1999})}\BibitemShut {NoStop}%
\bibitem [{\citenamefont {Rigol}\ \emph {et~al.}(2008)\citenamefont {Rigol},
  \citenamefont {Dunjko},\ and\ \citenamefont {Olshanii}}]{rigol_dunjko_08}%
  \BibitemOpen
  \bibfield  {author} {\bibinfo {author} {\bibfnamefont {M.}~\bibnamefont
  {Rigol}}, \bibinfo {author} {\bibfnamefont {V.}~\bibnamefont {Dunjko}}, \
  and\ \bibinfo {author} {\bibfnamefont {M.}~\bibnamefont {Olshanii}},\
  }\bibfield  {title} {\enquote {\bibinfo {title} {Thermalization and its
  mechanism for generic isolated quantum systems},}\ }\href@noop {} {\bibfield
  {journal} {\bibinfo  {journal} {Nature}\ }\textbf {\bibinfo {volume} {452}},\
  \bibinfo {pages} {854} (\bibinfo {year} {2008})}\BibitemShut {NoStop}%
\bibitem [{\citenamefont {Rigol}(2009{\natexlab{a}})}]{rigol_09a}%
  \BibitemOpen
  \bibfield  {author} {\bibinfo {author} {\bibfnamefont {M.}~\bibnamefont
  {Rigol}},\ }\bibfield  {title} {\enquote {\bibinfo {title} {Breakdown of
  thermalization in finite one-dimensional systems},}\ }\href@noop {}
  {\bibfield  {journal} {\bibinfo  {journal} {Phys. Rev. Lett.}\ }\textbf
  {\bibinfo {volume} {103}},\ \bibinfo {pages} {100403} (\bibinfo {year}
  {2009}{\natexlab{a}})}\BibitemShut {NoStop}%
\bibitem [{\citenamefont {Rigol}(2009{\natexlab{b}})}]{rigol_09b}%
  \BibitemOpen
  \bibfield  {author} {\bibinfo {author} {\bibfnamefont {M.}~\bibnamefont
  {Rigol}},\ }\bibfield  {title} {\enquote {\bibinfo {title} {Quantum quenches
  and thermalization in one-dimensional fermionic systems},}\ }\href@noop {}
  {\bibfield  {journal} {\bibinfo  {journal} {Phys. Rev. A}\ }\textbf {\bibinfo
  {volume} {80}},\ \bibinfo {pages} {053607} (\bibinfo {year}
  {2009}{\natexlab{b}})}\BibitemShut {NoStop}%
\bibitem [{\citenamefont {Santos}\ and\ \citenamefont
  {Rigol}(2010{\natexlab{a}})}]{santos_rigol_10b}%
  \BibitemOpen
  \bibfield  {author} {\bibinfo {author} {\bibfnamefont {L.~F.}\ \bibnamefont
  {Santos}}\ and\ \bibinfo {author} {\bibfnamefont {M.}~\bibnamefont {Rigol}},\
  }\bibfield  {title} {\enquote {\bibinfo {title} {Localization and the effects
  of symmetries in the thermalization properties of one-dimensional quantum
  systems},}\ }\href@noop {} {\bibfield  {journal} {\bibinfo  {journal} {Phys.
  Rev. E}\ }\textbf {\bibinfo {volume} {82}},\ \bibinfo {pages} {031130}
  (\bibinfo {year} {2010}{\natexlab{a}})}\BibitemShut {NoStop}%
\bibitem [{\citenamefont {Santos}\ and\ \citenamefont
  {Rigol}(2010{\natexlab{b}})}]{santos_rigol_10a}%
  \BibitemOpen
  \bibfield  {author} {\bibinfo {author} {\bibfnamefont {L.~F.}\ \bibnamefont
  {Santos}}\ and\ \bibinfo {author} {\bibfnamefont {M.}~\bibnamefont {Rigol}},\
  }\bibfield  {title} {\enquote {\bibinfo {title} {Onset of quantum chaos in
  one-dimensional bosonic and fermionic systems and its relation to
  thermalization},}\ }\href@noop {} {\bibfield  {journal} {\bibinfo  {journal}
  {Phys. Rev. E}\ }\textbf {\bibinfo {volume} {81}},\ \bibinfo {pages} {036206}
  (\bibinfo {year} {2010}{\natexlab{b}})}\BibitemShut {NoStop}%
\bibitem [{\citenamefont {Neuenhahn}\ and\ \citenamefont
  {Marquardt}(2012)}]{neuenhahn_marquardt_12}%
  \BibitemOpen
  \bibfield  {author} {\bibinfo {author} {\bibfnamefont {C.}~\bibnamefont
  {Neuenhahn}}\ and\ \bibinfo {author} {\bibfnamefont {F.}~\bibnamefont
  {Marquardt}},\ }\bibfield  {title} {\enquote {\bibinfo {title}
  {Thermalization of interacting fermions and delocalization in fock space},}\
  }\href@noop {} {\bibfield  {journal} {\bibinfo  {journal} {Phys. Rev. E}\
  }\textbf {\bibinfo {volume} {85}},\ \bibinfo {pages} {060101} (\bibinfo
  {year} {2012})}\BibitemShut {NoStop}%
\bibitem [{\citenamefont {Genway}\ \emph {et~al.}(2012)\citenamefont {Genway},
  \citenamefont {Ho},\ and\ \citenamefont {Lee}}]{genway_ho_12}%
  \BibitemOpen
  \bibfield  {author} {\bibinfo {author} {\bibfnamefont {S.}~\bibnamefont
  {Genway}}, \bibinfo {author} {\bibfnamefont {A.~F.}\ \bibnamefont {Ho}}, \
  and\ \bibinfo {author} {\bibfnamefont {D.~K.~K.}\ \bibnamefont {Lee}},\
  }\bibfield  {title} {\enquote {\bibinfo {title} {Thermalization of local
  observables in small hubbard lattices},}\ }\href@noop {} {\bibfield
  {journal} {\bibinfo  {journal} {Phys. Rev. A}\ }\textbf {\bibinfo {volume}
  {86}},\ \bibinfo {pages} {023609} (\bibinfo {year} {2012})}\BibitemShut
  {NoStop}%
\bibitem [{\citenamefont {Khatami}\ \emph {et~al.}(2013)\citenamefont
  {Khatami}, \citenamefont {Pupillo}, \citenamefont {Srednicki},\ and\
  \citenamefont {Rigol}}]{khatami_pupillo_13}%
  \BibitemOpen
  \bibfield  {author} {\bibinfo {author} {\bibfnamefont {E.}~\bibnamefont
  {Khatami}}, \bibinfo {author} {\bibfnamefont {G.}~\bibnamefont {Pupillo}},
  \bibinfo {author} {\bibfnamefont {M.}~\bibnamefont {Srednicki}}, \ and\
  \bibinfo {author} {\bibfnamefont {M.}~\bibnamefont {Rigol}},\ }\bibfield
  {title} {\enquote {\bibinfo {title} {Fluctuation-dissipation theorem in an
  isolated system of quantum dipolar bosons after a quench},}\ }\href@noop {}
  {\bibfield  {journal} {\bibinfo  {journal} {Phys. Rev. Lett.}\ }\textbf
  {\bibinfo {volume} {111}},\ \bibinfo {pages} {050403} (\bibinfo {year}
  {2013})}\BibitemShut {NoStop}%
\bibitem [{\citenamefont {Beugeling}\ \emph {et~al.}(2014)\citenamefont
  {Beugeling}, \citenamefont {Moessner},\ and\ \citenamefont
  {Haque}}]{beugeling_moessner_14}%
  \BibitemOpen
  \bibfield  {author} {\bibinfo {author} {\bibfnamefont {W.}~\bibnamefont
  {Beugeling}}, \bibinfo {author} {\bibfnamefont {R.}~\bibnamefont {Moessner}},
  \ and\ \bibinfo {author} {\bibfnamefont {M.}~\bibnamefont {Haque}},\
  }\bibfield  {title} {\enquote {\bibinfo {title} {Finite-size scaling of
  eigenstate thermalization},}\ }\href@noop {} {\bibfield  {journal} {\bibinfo
  {journal} {Phys. Rev. E}\ }\textbf {\bibinfo {volume} {89}},\ \bibinfo
  {pages} {042112} (\bibinfo {year} {2014})}\BibitemShut {NoStop}%
\bibitem [{\citenamefont {Kim}\ \emph {et~al.}(2014)\citenamefont {Kim},
  \citenamefont {Ikeda},\ and\ \citenamefont {Huse}}]{kim_14}%
  \BibitemOpen
  \bibfield  {author} {\bibinfo {author} {\bibfnamefont {H.}~\bibnamefont
  {Kim}}, \bibinfo {author} {\bibfnamefont {T.~N.}\ \bibnamefont {Ikeda}}, \
  and\ \bibinfo {author} {\bibfnamefont {D.~A.}\ \bibnamefont {Huse}},\
  }\bibfield  {title} {\enquote {\bibinfo {title} {Testing whether all
  eigenstates obey the eigenstate thermalization hypothesis},}\ }\href@noop {}
  {\bibfield  {journal} {\bibinfo  {journal} {Phys. Rev. E}\ }\textbf {\bibinfo
  {volume} {90}},\ \bibinfo {pages} {052105} (\bibinfo {year}
  {2014})}\BibitemShut {NoStop}%
\bibitem [{\citenamefont {Sorg}\ \emph {et~al.}(2014)\citenamefont {Sorg},
  \citenamefont {Vidmar}, \citenamefont {Pollet},\ and\ \citenamefont
  {Heidrich-Meisner}}]{sorg_vidmar_14}%
  \BibitemOpen
  \bibfield  {author} {\bibinfo {author} {\bibfnamefont {S.}~\bibnamefont
  {Sorg}}, \bibinfo {author} {\bibfnamefont {L.}~\bibnamefont {Vidmar}},
  \bibinfo {author} {\bibfnamefont {L.}~\bibnamefont {Pollet}}, \ and\ \bibinfo
  {author} {\bibfnamefont {F.}~\bibnamefont {Heidrich-Meisner}},\ }\bibfield
  {title} {\enquote {\bibinfo {title} {Relaxation and thermalization in the
  one-dimensional {Bose-Hubbard} model: A case study for the interaction
  quantum quench from the atomic limit},}\ }\href@noop {} {\bibfield  {journal}
  {\bibinfo  {journal} {Phys. Rev. A}\ }\textbf {\bibinfo {volume} {90}},\
  \bibinfo {pages} {033606} (\bibinfo {year} {2014})}\BibitemShut {NoStop}%
\bibitem [{\citenamefont {Rigol}\ and\ \citenamefont
  {Srednicki}(2012)}]{AETRigol}%
  \BibitemOpen
  \bibfield  {author} {\bibinfo {author} {\bibfnamefont {M.}~\bibnamefont
  {Rigol}}\ and\ \bibinfo {author} {\bibfnamefont {M.}~\bibnamefont
  {Srednicki}},\ }\bibfield  {title} {\enquote {\bibinfo {title} {Alternatives
  to eigenstate thermalization},}\ }\href@noop {} {\bibfield  {journal}
  {\bibinfo  {journal} {Phys. Rev. Lett.}\ }\textbf {\bibinfo {volume} {108}},\
  \bibinfo {pages} {110601} (\bibinfo {year} {2012})}\BibitemShut {NoStop}%
\bibitem [{\citenamefont {Ikeda}\ \emph {et~al.}(2013)\citenamefont {Ikeda},
  \citenamefont {Watanabe},\ and\ \citenamefont {Ueda}}]{FSSIkeda}%
  \BibitemOpen
  \bibfield  {author} {\bibinfo {author} {\bibfnamefont {T.~N.}\ \bibnamefont
  {Ikeda}}, \bibinfo {author} {\bibfnamefont {Y.}~\bibnamefont {Watanabe}}, \
  and\ \bibinfo {author} {\bibfnamefont {M.}~\bibnamefont {Ueda}},\ }\bibfield
  {title} {\enquote {\bibinfo {title} {Finite-size scaling analysis of the
  eigenstate thermalization hypothesis in a one-dimensional interacting bose
  gas},}\ }\href@noop {} {\bibfield  {journal} {\bibinfo  {journal} {Phys. Rev.
  E}\ }\textbf {\bibinfo {volume} {87}},\ \bibinfo {pages} {012125} (\bibinfo
  {year} {2013})}\BibitemShut {NoStop}%
\bibitem [{\citenamefont {Steinigeweg}\ \emph {et~al.}(2014)\citenamefont
  {Steinigeweg}, \citenamefont {Khodja}, \citenamefont {Niemeyer},
  \citenamefont {Gogolin},\ and\ \citenamefont {Gemmer}}]{PLETHSteinigweg}%
  \BibitemOpen
  \bibfield  {author} {\bibinfo {author} {\bibfnamefont {R.}~\bibnamefont
  {Steinigeweg}}, \bibinfo {author} {\bibfnamefont {A.}~\bibnamefont {Khodja}},
  \bibinfo {author} {\bibfnamefont {H.}~\bibnamefont {Niemeyer}}, \bibinfo
  {author} {\bibfnamefont {C.}~\bibnamefont {Gogolin}}, \ and\ \bibinfo
  {author} {\bibfnamefont {J.}~\bibnamefont {Gemmer}},\ }\bibfield  {title}
  {\enquote {\bibinfo {title} {Pushing the limits of the eigenstate
  thermalization hypothesis towards mesoscopic quantum systems},}\ }\href@noop
  {} {\bibfield  {journal} {\bibinfo  {journal} {Phys. Rev. Lett.}\ }\textbf
  {\bibinfo {volume} {112}},\ \bibinfo {pages} {130403} (\bibinfo {year}
  {2014})}\BibitemShut {NoStop}%
\bibitem [{\citenamefont {Khlebnikov}\ and\ \citenamefont
  {Kruczenski}(2013)}]{Khlebnikov2013yia}%
  \BibitemOpen
  \bibfield  {author} {\bibinfo {author} {\bibfnamefont {S.}~\bibnamefont
  {Khlebnikov}}\ and\ \bibinfo {author} {\bibfnamefont {M.}~\bibnamefont
  {Kruczenski}},\ }\bibfield  {title} {\enquote {\bibinfo {title}
  {Thermalization of isolated quantum systems},}\ }\href@noop {} {\bibfield
  {journal} {\bibinfo  {journal} {arXiv:1312.4612 [cond-mat.stat-mech]}\ }
  (\bibinfo {year} {2013})}\BibitemShut {NoStop}%
\bibitem [{\citenamefont {Garrison}\ and\ \citenamefont
  {Grover}(2018)}]{Garrison2015lva}%
  \BibitemOpen
  \bibfield  {author} {\bibinfo {author} {\bibfnamefont {James~R.}\
  \bibnamefont {Garrison}}\ and\ \bibinfo {author} {\bibfnamefont {Tarun}\
  \bibnamefont {Grover}},\ }\bibfield  {title} {\enquote {\bibinfo {title}
  {Does a single eigenstate encode the full hamiltonian?}}\ }\href {\doibase
  10.1103/PhysRevX.8.021026} {\bibfield  {journal} {\bibinfo  {journal} {Phys.
  Rev. X}\ }\textbf {\bibinfo {volume} {8}},\ \bibinfo {pages} {021026}
  (\bibinfo {year} {2018})}\BibitemShut {NoStop}%
\bibitem [{\citenamefont {Polkovnikov}\ \emph {et~al.}(2011)\citenamefont
  {Polkovnikov}, \citenamefont {Sengupta}, \citenamefont {Silva},\ and\
  \citenamefont {Vengalattore}}]{NDPolkovnikov}%
  \BibitemOpen
  \bibfield  {author} {\bibinfo {author} {\bibfnamefont {A.}~\bibnamefont
  {Polkovnikov}}, \bibinfo {author} {\bibfnamefont {K.}~\bibnamefont
  {Sengupta}}, \bibinfo {author} {\bibfnamefont {A.}~\bibnamefont {Silva}}, \
  and\ \bibinfo {author} {\bibfnamefont {M.}~\bibnamefont {Vengalattore}},\
  }\bibfield  {title} {\enquote {\bibinfo {title} {\textit{Colloquium} :
  Nonequilibrium dynamics of closed interacting quantum systems},}\ }\href@noop
  {} {\bibfield  {journal} {\bibinfo  {journal} {Rev. Mod. Phys.}\ }\textbf
  {\bibinfo {volume} {83}},\ \bibinfo {pages} {863--883} (\bibinfo {year}
  {2011})}\BibitemShut {NoStop}%
\bibitem [{\citenamefont {D'Alessio}\ \emph {et~al.}(2015)\citenamefont
  {D'Alessio}, \citenamefont {Kafri}, \citenamefont {Polkovnikov},\ and\
  \citenamefont {Rigol}}]{mrigolreview2015}%
  \BibitemOpen
  \bibfield  {author} {\bibinfo {author} {\bibfnamefont {L.}~\bibnamefont
  {D'Alessio}}, \bibinfo {author} {\bibfnamefont {Y.}~\bibnamefont {Kafri}},
  \bibinfo {author} {\bibfnamefont {A.}~\bibnamefont {Polkovnikov}}, \ and\
  \bibinfo {author} {\bibfnamefont {M.}~\bibnamefont {Rigol}},\ }\bibfield
  {title} {\enquote {\bibinfo {title} {From quantum chaos and eigenstate
  thermalization to statistical mechanics and thermodynamics},}\ }\href@noop {}
  {\bibfield  {journal} {\bibinfo  {journal} {arXiv:1509.06411
  [cond-mat.stat-mech]}\ } (\bibinfo {year} {2015})}\BibitemShut {NoStop}%
\bibitem [{\citenamefont {{Qi}}\ and\ \citenamefont
  {{Ranard}}(2017)}]{newStanford}%
  \BibitemOpen
  \bibfield  {author} {\bibinfo {author} {\bibfnamefont {X.-L.}\ \bibnamefont
  {{Qi}}}\ and\ \bibinfo {author} {\bibfnamefont {D.}~\bibnamefont
  {{Ranard}}},\ }\bibfield  {title} {\enquote {\bibinfo {title} {{Determining a
  local Hamiltonian from a single eigenstate}},}\ }\href@noop {} {\bibfield
  {journal} {\bibinfo  {journal} {ArXiv e-prints}\ } (\bibinfo {year}
  {2017})},\ \Eprint {http://arxiv.org/abs/1712.01850} {arXiv:1712.01850
  [quant-ph]} \BibitemShut {NoStop}%
\bibitem [{\citenamefont {Fratus}\ and\ \citenamefont
  {Srednicki}(2015)}]{myPaperOne}%
  \BibitemOpen
  \bibfield  {author} {\bibinfo {author} {\bibfnamefont {Keith~R.}\
  \bibnamefont {Fratus}}\ and\ \bibinfo {author} {\bibfnamefont {Mark}\
  \bibnamefont {Srednicki}},\ }\bibfield  {title} {\enquote {\bibinfo {title}
  {Eigenstate thermalization in systems with spontaneously broken symmetry},}\
  }\href {\doibase 10.1103/PhysRevE.92.040103} {\bibfield  {journal} {\bibinfo
  {journal} {Phys. Rev. E}\ }\textbf {\bibinfo {volume} {92}},\ \bibinfo
  {pages} {040103} (\bibinfo {year} {2015})}\BibitemShut {NoStop}%
\bibitem [{\citenamefont {Mondaini}\ \emph {et~al.}(2016)\citenamefont
  {Mondaini}, \citenamefont {Fratus}, \citenamefont {Srednicki},\ and\
  \citenamefont {Rigol}}]{myPaperTwo}%
  \BibitemOpen
  \bibfield  {author} {\bibinfo {author} {\bibfnamefont {Rubem}\ \bibnamefont
  {Mondaini}}, \bibinfo {author} {\bibfnamefont {Keith~R.}\ \bibnamefont
  {Fratus}}, \bibinfo {author} {\bibfnamefont {Mark}\ \bibnamefont
  {Srednicki}}, \ and\ \bibinfo {author} {\bibfnamefont {Marcos}\ \bibnamefont
  {Rigol}},\ }\bibfield  {title} {\enquote {\bibinfo {title} {Eigenstate
  thermalization in the two-dimensional transverse field ising model},}\ }\href
  {\doibase 10.1103/PhysRevE.93.032104} {\bibfield  {journal} {\bibinfo
  {journal} {Phys. Rev. E}\ }\textbf {\bibinfo {volume} {93}},\ \bibinfo
  {pages} {032104} (\bibinfo {year} {2016})}\BibitemShut {NoStop}%
\bibitem [{\citenamefont {{Fratus}}\ and\ \citenamefont
  {{Srednicki}}(2016)}]{myPaperThree}%
  \BibitemOpen
  \bibfield  {author} {\bibinfo {author} {\bibfnamefont {K.~R.}\ \bibnamefont
  {{Fratus}}}\ and\ \bibinfo {author} {\bibfnamefont {M.}~\bibnamefont
  {{Srednicki}}},\ }\bibfield  {title} {\enquote {\bibinfo {title} {{Eigenstate
  Thermalization and Spontaneous Symmetry Breaking in the One-Dimensional
  Transverse-Field Ising Model with Power-Law Interactions}},}\ }\href@noop {}
  {\bibfield  {journal} {\bibinfo  {journal} {ArXiv e-prints}\ } (\bibinfo
  {year} {2016})},\ \Eprint {http://arxiv.org/abs/1611.03992} {arXiv:1611.03992
  [cond-mat.stat-mech]} \BibitemShut {NoStop}%
\bibitem [{\citenamefont {Cardy}(2012)}]{cardy2012finite}%
  \BibitemOpen
  \bibfield  {author} {\bibinfo {author} {\bibfnamefont {John}\ \bibnamefont
  {Cardy}},\ }\href@noop {} {\emph {\bibinfo {title} {Finite-size scaling}}},\
  Vol.~\bibinfo {volume} {2}\ (\bibinfo  {publisher} {Elsevier},\ \bibinfo
  {year} {2012})\BibitemShut {NoStop}%
\bibitem [{\citenamefont {Binder}(1981)}]{Binder81}%
  \BibitemOpen
  \bibfield  {author} {\bibinfo {author} {\bibfnamefont {K.}~\bibnamefont
  {Binder}},\ }\bibfield  {title} {\enquote {\bibinfo {title} {Critical
  properties from monte carlo coarse graining and renormalization},}\
  }\href@noop {} {\bibfield  {journal} {\bibinfo  {journal} {Phys. Rev. Lett.}\
  }\textbf {\bibinfo {volume} {47}},\ \bibinfo {pages} {693--696} (\bibinfo
  {year} {1981})}\BibitemShut {NoStop}%
\bibitem [{\citenamefont {Dyson}(1969)}]{Dyson1969}%
  \BibitemOpen
  \bibfield  {author} {\bibinfo {author} {\bibfnamefont {F.~J.}\ \bibnamefont
  {Dyson}},\ }\bibfield  {title} {\enquote {\bibinfo {title} {Existence of a
  phase-transition in a one-dimensional ising ferromagnet},}\ }\href@noop {}
  {\bibfield  {journal} {\bibinfo  {journal} {Communications in Mathematical
  Physics}\ }\textbf {\bibinfo {volume} {12}},\ \bibinfo {pages} {91--107}
  (\bibinfo {year} {1969})}\BibitemShut {NoStop}%
\bibitem [{Note1()}]{Note1}%
  \BibitemOpen
  \bibinfo {note} {See Supplemental Material at [URL will be inserted by
  publisher] for a more detailed explanation of the precise numerical
  extraction procedure we have used, which includes a discussion of how well
  the subsystem Hamiltonian agrees with the exact Hamiltonian, how we define
  the temperature of an eigenstate, and how well our results agree from one
  choice of eigenstate to another.}\BibitemShut {Stop}%
\end{thebibliography}%

\end{document}